\documentclass{article} 
\usepackage{iclr2020_conference,times}

\usepackage{amsmath,amsfonts,bm}

\def\eqref#1{equation~\ref{#1}}

\def\1{\bm{1}}

\DeclareMathAlphabet{\mathsfit}{\encodingdefault}{\sfdefault}{m}{sl}
\SetMathAlphabet{\mathsfit}{bold}{\encodingdefault}{\sfdefault}{bx}{n}

\usepackage{hyperref}
\usepackage{url}
\usepackage{graphicx}
\usepackage{subfigure}
\usepackage{amsmath}
\usepackage{amsfonts}
\usepackage{marvosym}

\title{Stereo Endoscopic Image Super-Resolution Using Disparity-Constrained Parallel \\ Attention}

\iclrfinalcopy
\author{Tianyi Zhang\textsuperscript{1,2}, Yun Gu\textsuperscript{1,2},
	Xiaolin Huang\textsuperscript{1,2}, Enmei Tu\textsuperscript{1} \& Jie Yang\textsuperscript{1,2,\Letter} \\
	Institute of Image Processing and Pattern Recognition\textsuperscript{1},
	Institute of Medical Robotics\textsuperscript{2}\\
	Shanghai Jiao Tong University\\
	Shanghai, China \\
	\texttt{\{autozty,geron762,xiaolinhuang,tuen,jieyang\}@sjtu.edu.cn} 
}

\begin{document}

\maketitle

\begin{abstract}
With the popularity of stereo cameras in computer assisted surgery techniques, a second viewpoint would provide additional information in surgery. However, how to effectively access and use stereo information for the super-resolution (SR) purpose is often a challenge.
In this paper, we propose a disparity-constrained stereo super-resolution network (DCSSRnet) to simultaneously compute a super-resolved image in a stereo image pair. In particular, we incorporate a disparity-based constraint mechanism into the generation of SR images in a deep neural network framework with an additional atrous parallax-attention modules.
Experiment results on laparoscopic images demonstrate that the proposed framework outperforms current SR methods on both quantitative and qualitative evaluations.
Our DCSSRnet provides a promising solution on enhancing spatial resolution of stereo image pairs, which will be extremely beneficial for the endoscopic surgery. 
\end{abstract}

\section{Introduction}
\label{intro}
\begin{figure}[t]
	\centering
	\includegraphics[scale=0.15]{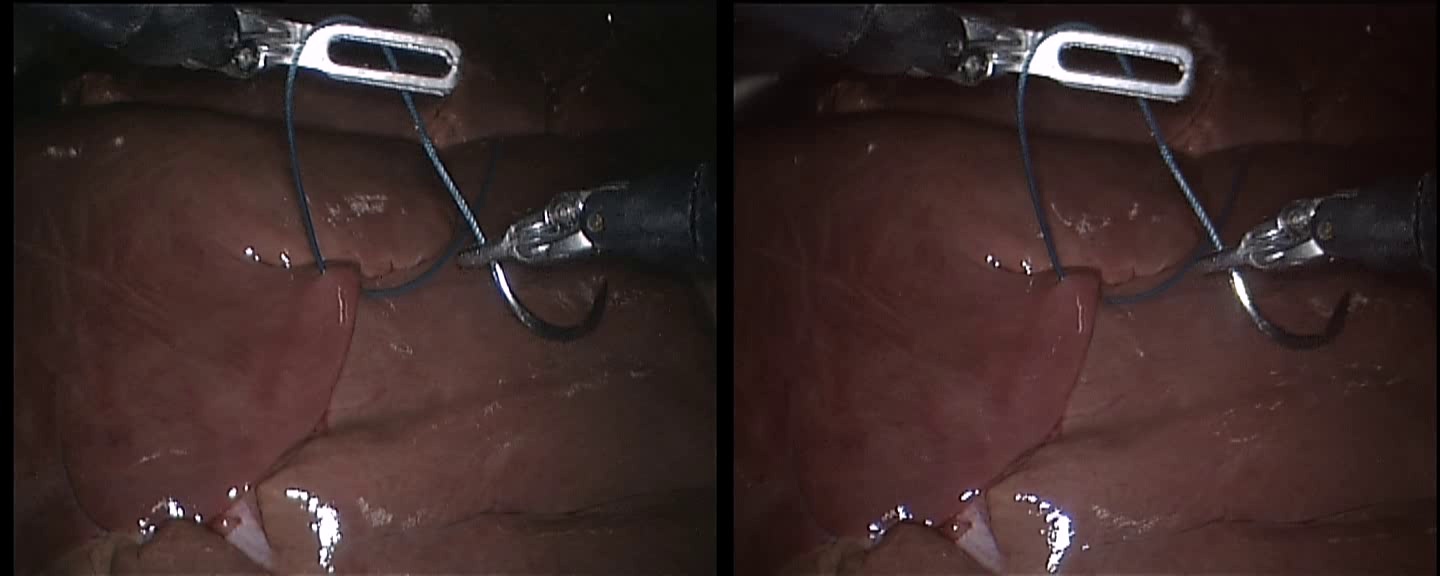}
	\caption{Laparoscopic images taken by stereo cameras.}
	\label{fig1}
\end{figure}

With the development of optical and electronic engineering, the endoscope system has become an essential component of the interventional surgery; i.e., when a biopsy is performed on patients, the physicians can greatly benefit from the real-time endoscopic images. Though the majority of the current endoscope systems are monoscopic, stereo cameras have received attention in computer assisted surgery. Fig.~\ref{fig1} presents the da Vinci surgical robotic system equipped with laparoscopic stereo cameras for robot-assisted surgery. 

However, obtaining high-quality images suffers from the challenges of limited size of optical sensors, the conned space and thin vessels. In order to achieve a high resolution image containing sharp texture and edge details, the super-resolution (SR) technique can be used to recover high-resolution (HR) images from its low-resolution (LR) ones. 

In recent years, deep-learning-based SR algorithms have shown promising performance. ~\citet{r2} propose an end-to-end mapping by using a three-layer convolutional neural network (SRCNN). ~\citet{r3} present a very deep convolutional network with residual structure (VDSR). ~\citet{r14} propose a deep recursive residual network (DRRN) to more effectively generate the feature maps in SR problem. In addition, the success of SR-based techniques has inspired several improvements on medical imaging techniques, such as using SR method on magnetic resonance images~\citep{r5,r6} and microscopic images~\citep{r9}.

Recently, there have been notable works that utilized the meaningful information in the stereo images in order to improve the results. Jeon et al. proposed a stereo-enhancement super-resolution network (StereoSR)~\citep{r7} to extract information from a base image and then stack to an auxiliary image in order to generate a more accurate result. However, this technique is not suitable for stereo images that contains variant disparity.
~\citet{r13} proposed a parallax-attention stereo superresolution network (PASSR) that  incorporate the stereo correspondence into the SR technique. The residual atrous spatial pyramid pooling (ASPP) module is first used to generate feature maps. Further, the disparity mask with the stereo correspondence will be achieved from feeding extracted features into a parallax-attention module (PAM).
However, the consistency of disparity information between LR and HR pairs are out of consideration in above methods. Moreover, only using the epipolar line in parallax-attention mechanism based stereo image pairs has received limited attention. Pixels near the epipolar line can also affect the attention map, since the stereo pair may be not accurately registered.

In this paper, we propose a disparity-constrained stereo super-resolution network (DCSSR) to incorporate the disparity constraint between low and high resolution. A symmetric network architecture is proposed to simultaneously generate left and right SR images. In addition, an atrous parallax-attention module is proposed to generate LR and SR disparity masks. Taking the difference of LR and HR disparity masks as a regularization, there would be the same stereo correspondence of the super-resolved results as the LR images.

Experiment results on laparoscopic images reveal that the proposed architecture outperforms other SR methods on quantitative measurements. 
Results indicate that the SR images recovered by proposed method are more subjectively realistic.

\section{Method}
\label{sec:2}
\subsection{Workflow}

\begin{figure}[htbp]
	\centering
	\includegraphics[scale=0.4]{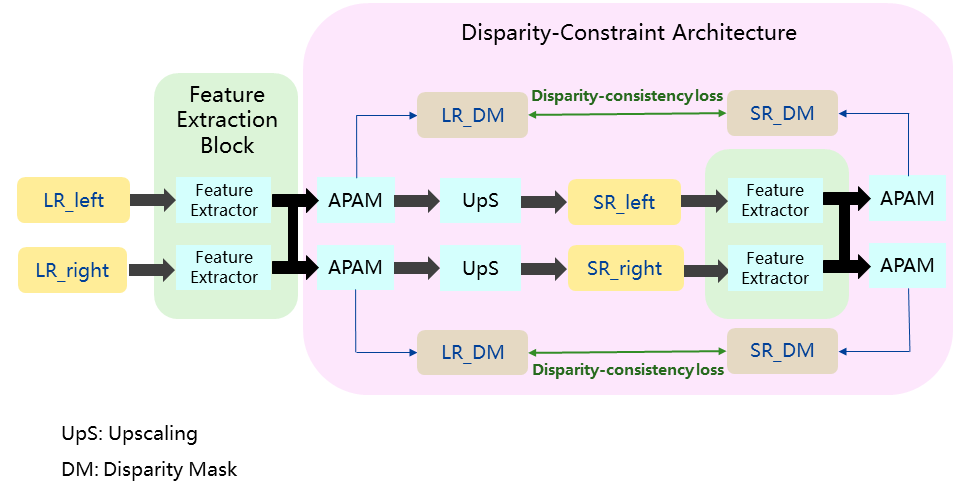}
	\caption{The proposed network architecture. The yellow boxes respectively represent the input LR images and output SR images.}
	\label{fig_net}
\end{figure}

As shown in Fig.\ref{fig_net}, the workflow of the proposed method can be divided into the feature extraction block and the disparity-constraint architecture.


\subsection{Feature Extraction Block}
\begin{figure}[htbp]
	\centering
	\includegraphics[scale=0.4]{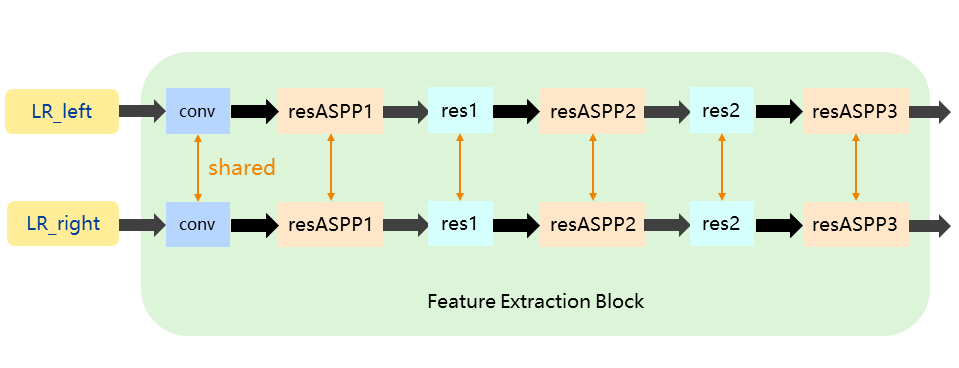}
	\caption{The architecture of the feature extraction block. The "res" module means residual block, which contains two convolutional layers and the skip connection between head and tail. Details of the "resASPP" module are described by~\citet{r13}.}
	\label{figFE}
\end{figure}

Before learning stereo correspondences, large receptive field and multi-scale feature learning are benefitial for feature representation. In DCSSRnet, we use the residual ASPP module~\citep{r13} as the backbone of feature extractor, which constructed by alternately cascading a residual ASPP block
with a residual block. As illustrated in Fig.~\ref{figFE}, three residual block and two residual ASPP block share the weights in the process of extracting left and right features. In order to enlarge the receptive field, three dilated convolutions with rates of 1, 4, 8 are conducted in each resASPP block, and the results are concatenated and fused by
a $1\times1$ convolution.
\subsection{Disparity-Constraint Architecture}
In order to preserve the stereo correspondence between left and right images in the original images and its results, we develop a disparity-constraint architecture by utilizing the left-right consistency in LR and SR image pairs. 

For left and right feature maps, ~\citet{r13} proposed PAM to capture global correspondence. In PAM, features along the epipolar line are utilized to build sparse attention masks by batch-wise matrix multiplication. Different from PAM~\citep{r13}, our Atrous PAM (APAM) utilizes multiple-line similar features rather than the epipolar line to generate disparity masks. Since a pixel in left image is corresponding to a cluster of pixels around the epipolar line in right image, it is not convenient to only use the epipolar line for the stereo image alignment. In APAM, we use residual atrous spatial pyramid pooling to enlarge the receptive field and fuse variant multiple-line features. Then the batch-wise matrix multiplication is applied between left and right feature maps, followed by a softmax layer to generate the disparity mask. More details around the implementation of PAM are described by~\citet{r13}.

The generated disparity mask is a reflection of stereo correspondence between left and right images. For the purpose of acquiring more realistic results, there must be almost consistent disparity mask between the super-resolved image pairs and the original ones.  In our disparity-constraint architecture, the LR and SR disparity masks are both generated in training stage of network. As a regularization, a loss function is proposed to minimize the trilinear-interpolated LR disparity mask and the SR one.

\subsection{Loss Function}
For the training of the network, we design a composite loss function, which consists of three parts, including MSE loss, disparity-consistency loss, and APAM loss. The loss function can be presented as

\begin{equation}\label{eq6}
\centering
\min \limits_{\theta}{\ \frac{1}{n}\sum_{i=1}^n (L_{MSE} + \alpha (L_{DC} + L_{APAM}))}
\end{equation}
where $n$ is the batch size and $\alpha$ is a factor to adjust the proportion of each part. 

The mean square error (MSE) loss measures the similarity of image intensity between reconstructed SR image and the HR image. Taken the left image as an example, the MSE loss can be presented as
\begin{equation}
\centering
L_{MSE} = \|I^{SR}_l - I^{HR}_l\|_2^2
\end{equation}
where $I^{SR}_l$ and $I^{HR}_l$ respectively represent the SR result and HR image. 

Considering the consistency between LR and SR disparity masks which are respectively generated by LR and SR image pairs, we design a disparity-consistency loss function as
\begin{equation}
\centering
L_{DC} = \|ups(M_{l\to r}^{LR}) - M_{l\to r}^{SR}\|_2^2 
+ \|ups(M_{r\to l}^{LR}) - M_{r\to l}^{SR}\|_2^2
\end{equation}
where the $ups()$ represents the upscaling function to keep the same size of disparity masks between $M^{LR}$ and $M^{SR}$.


Similar to PAM loss~\citep{r13}, we use APAM loss to regularize original LR disparity masks. The APAM loss consists of photometric loss, smoothness loss and cycle loss, which is presented as (\ref{eqapam}).
\begin{equation}\label{eqapam}
\centering
L_{APAM} = L_{photo} + L_{smooth(l\to r)} + L_{smooth_(r\to l)} + L_{cycle}
\end{equation}
Since the disparity mask is available in APAM, the right image can be generated by multiplying the disparity mask to the left image matrix. The photometric loss can be formulated as
\begin{equation}\label{eqphoto}
\centering
L_{photo} = \|I_l^{LR} - (M_{r\to l}^{LR}\otimes I_r^{LR})\|_1 + \|I_r^{LR} - (M_{l\to r}^{LR}\otimes I_l^{LR})\|_1
\end{equation}
where $\otimes$ denotes batch-wise matrix multiplication. Additionally, a smoothness loss function is introduced to generate smoother and consistent disparity mask:
\begin{equation}\label{eqsmooth}
\centering
\begin{aligned}
L_{smooth(l\to r)} &= \sum_{i,j,k} (\|M_{l\to r}^{LR}(i,j,k) -
M_{l\to r}^{LR}(i+1,j,k)\|_1 \\
&+
\|M_{l\to r}^{LR}(i,j,k) -
M_{l\to r}^{LR}(i,j+1,k+1)\|_1
\end{aligned}
\end{equation}
Also, a cycle loss is utilized to keep the cycle consistency of disparity mask. The cycle loss function can be presented as
\begin{equation}\label{eqcycle}
\centering
L_{cycle} = \|M_{l\to r}^{LR}\otimes
M_{r\to l}^{LR} - \textbf{I}\|_1 + \|M_{r\to l}^{LR}\otimes
M_{l\to r}^{LR} - \textbf{I}\|_1 
\end{equation}
where $\textbf{I}\in \mathbb{R}^{H\times W\times W} $ denotes a stack of $H$ identity matrices.

\section{Experiments and Results}
\subsection{Dataset}
We used 4560 high-resolution (512$\times$ 512) stereo image pairs that are collected by stereo camera of the da Vanci surgical system. Similar to ~\citet{r13}, stereo HR frames in two videos are downsampled with the scale factor $2\times$ and $4\times$, and cropped to $30 \times 90$ patches with a stride of 20 pixels. Data augmentation has been used by flipping along the x- and y-axis, and random crops. For the experiments, 870 frames are used to validate the model, and 870 frames in another video collected by the same device are used for testing.

For comparison, we also used the same dataset in other learning-based methods, including SRCNN~\citep{r2}, VDSR~\citep{r3}, DRRN~\citep{r14}, stereoSR~\citep{r7} and PASSR~\citep{r13}. The codes and training measures provided by the authors of these methods were utilized to conduct experiments.
\subsection{Training Details}
The model has been implemented in Pytorch, using an Nvidia GTX 1080 Ti GPU. The convolutional filters are initialized by Xavier initializer~\citep{r8}. The Adam optimizer is used for learning the model and its momentum is set to 0.9; the initial learning rate is set to  $2\times10^{-4}$, and would be reduced to half after every 30 epochs. The network has been trained for 80 epochs with the batch size of 8 and 4 for $2\times$ and $4\times$. We also set the factor $\alpha$ to 0.005 in order to rebalance the Eq.~(\ref{eq6}). For the disparity constraint loss function, trilinear interpolation is used in the upsampling of LR disparity masks.

\subsection{Quantitative Evaluation}
The Peak Signal-to-Noise Ratio (PSNR) is a broadly used quantitative measurement in image super-resolution. The SSIM is also quantified as a perceptual metric of image affinity. We compare our algorithm with several SR methods that are based on deep neural network with similar structures. The quantitative results are shown in Table~\ref{tab1}. By taking the similarity between stereo images into consideration, our proposed algorithm performs better than other networks on PSNR and SSIM.
\begin{table}
	\caption{Performance comparison between current methods and proposed method.}\label{tab1}\centering
	\begin{tabular}{c|c|c||c|c}
		\hline
		Methods & PSNR($\times$2) & SSIM($\times$2) & PSNR($\times$4) &	SSIM($\times$4)\\
		\hline
		bicubic & 35.019 & 0.9861 & 32.819 & 0.9694\\
		SRCNN & 39.980 & 0.9939 & 33.974 & 0.9785 \\
		VDSR & 40.706 & 0.9946 & 34.696 & 0.9796\\
		DRRN & 40.975 & 0.9943 & 35.458 & 0.9804 \\
		StereoSR & 41.204 & 0.9947 & - & - \\
		PASSR & 43.018 & 0.9955  & 35.092 & 0.9792\\
		DCSSR(proposed) & \textbf{43.610} & \textbf{0.9957} & \textbf{35.598} & \textbf{0.9809} \\
		\hline

	\end{tabular}
\end{table}

\subsection{Qualitative Evaluation}
\begin{figure}[t]
	\centering
	\subfigure[HR image]
	{
		\includegraphics[scale=0.3]{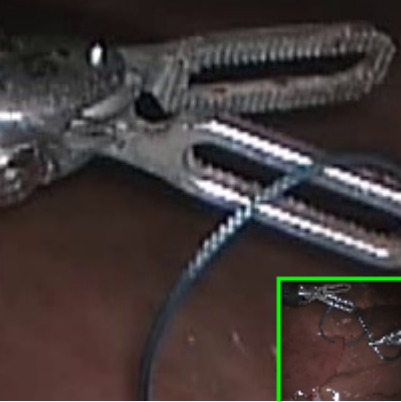}\vspace{3.pt}
	}
	\subfigure[bicubic]
	{
		\includegraphics[scale=0.3]{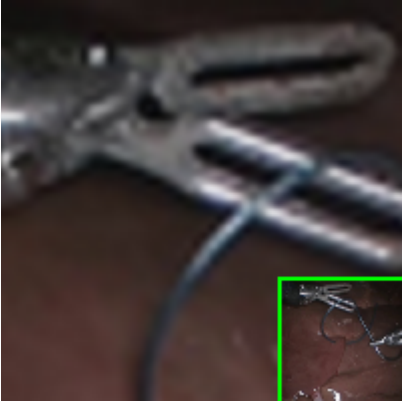}\vspace{3.pt}
	}
	\subfigure[SRCNN]
	{
		\includegraphics[scale=0.3]{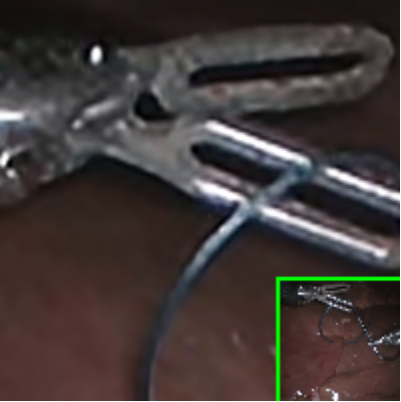}\vspace{3.pt}
	}
	\subfigure[VDSR]
	{
		\includegraphics[scale=0.3]{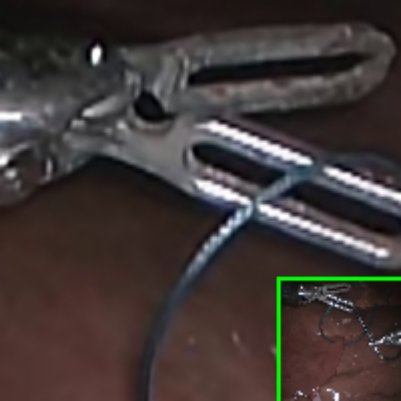}\vspace{3.pt}
	}
	\subfigure[DRRN]
	{
		\includegraphics[scale=0.3]{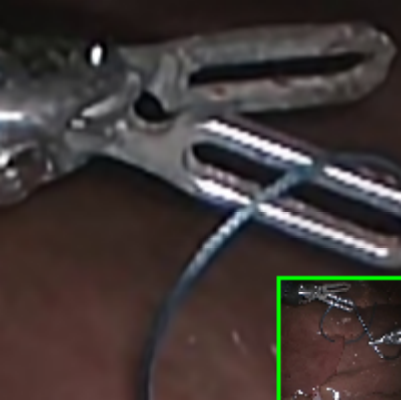}\vspace{3.pt}
	}
	\subfigure[PASSR]
	{
		\includegraphics[scale=0.3]{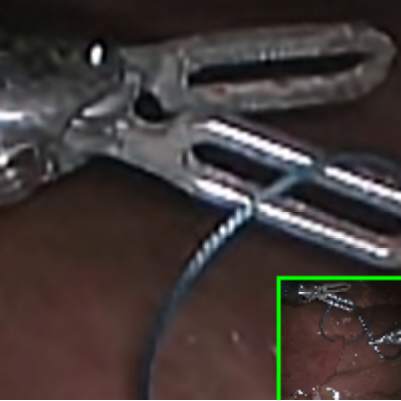}\vspace{3.pt}
	}
	\subfigure[\textbf{DCSSR}]
	{
		\includegraphics[scale=0.3]{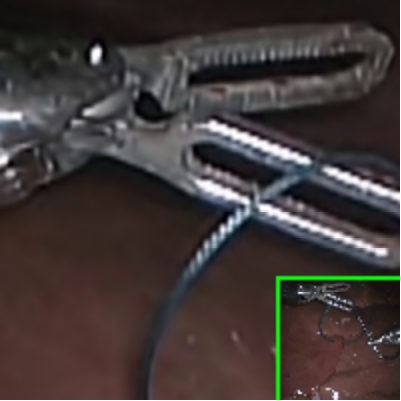}\vspace{3.pt}
	}
	\centering
	\caption{Super-resolution laparascopic images recovered by bicubic interpolation, SRCNN, VDSR, DRRN, PASSRnet, our method and corresponding HR image with scale factor 4×.}
	\label{fig2}
\end{figure}

Fig.~\ref{fig2} presents the SR performance and comparison with other techniques. Details can be seen with zooming-in on the regions while the green box in the lower right
corner shows the whole SR image. The results convey that, single image SR methods provide less details compared to stereo approaches. As compared to the PASSR, our DCSSRnet uses disparity-constrained correspondence in stereo image pairs to improve the SR performance of edge and texture details, as
shown in Fig.~\ref{fig2}.

However, there still exist several cases in which artifacts occur only in the left image. It is a problem for our method to generate a quite accurate disparity mask in this case. As a result, our algorithm has limitations when the left and right images are inconsistent to some extent.

\section{Conclusion}
In this work, we propose a new super-resolution method for stereo endoscopic images using a symmetric disparity-constrained neural network. In particular, we improve the SR algorithms in (1) disparity-constraint regularization, (2) symmetric architecture and (3) a feasible method of stereo alignment. Moreover, the proposed model is able to preserve the original structure of super-resolved stereo images by effectively utilizes the disparity consistency of SR and LR stereo image pairs. We have observed from our experiments that our method is more effective due to the constrained disparity in different resolutions. We believe that this method can be used in surgery to improve the imaging quality of stereo cameras and motivate fast deep models for super-resolution of stereo videos.

\subsubsection*{Acknowledgments}
The This research is partly supported by Ministry of Science and Technology, China\\
(No. 2019YFB1311503), and Committee of Science and Technology, Shanghai, China (No. 19510711200).

\bibliography{iclr2020_conference}
\bibliographystyle{iclr2020_conference}

\end{document}